\documentclass[aps,prl,reprint,superscriptaddress]{revtex4-2}

\usepackage[T1]{fontenc}
\usepackage[utf8]{inputenc}
\usepackage{bm,amsmath,amssymb,mathtools}
\usepackage{graphicx}
\usepackage[dvipsnames]{xcolor}
\usepackage{times}
\usepackage{microtype}
\usepackage[colorlinks=true,allcolors=blue]{hyperref}

\usepackage[capitalise]{cleveref}
\usepackage[normalem]{ulem}

\DeclareMathOperator*{\argmax}{argmax}
\DeclareMathOperator{\sgn}{sgn}

\allowdisplaybreaks[2]

\begin{document}

\title{Optimal Navigation in Stochastic and Disordered Gridworlds}

\author{K\'evin Bila\"i Biloa}
\author{Olivier Pierre-Louis}
\affiliation{Universit\'e Lyon 1, CNRS, Institut Lumi\`ere Mati\`ere, UMR5306  69622 Villeurbanne, France}

\date{\today}

\begin{abstract}
Navigation in complex and noisy environments is a key issue in diverse fields from biology to engineering.
Despite extensive progress in numerical optimization methods for computing navigation policies, 
insights into how disorder reshapes optimal navigation remain elusive. 
To address this question, we investigate the navigation of a Brownian particle in a disordered energy landscape, 
modeled as a lattice with randomly distributed traps.
Using dynamic programming, we compute the optimal navigation policies that minimize the mean first-passage time to a target site.
To quantify the impact of disorder, we introduce a density of change from a Kullback–Leibler 
divergence, which captures how the optimal policy is reshaped by either the presence of disorder or the knowledge of its configuration.
Our results reveal a non-monotonic dependence of the change of the policy on trap concentration, with a pronounced maximum.
In the fluctuation-dominated regime where the navigation bias is weak, we derive an analytical expression for the density of change, 
and demonstrate that the maximum occurs unexpectedly at low trap concentrations.
\end{abstract}

\maketitle

\paragraph{Introduction---}
Navigation is a vital challenge for living organisms during foraging and mating~\cite{Vickers2000,montello2005navigation,hoinville2018optimal}, 
and is also essential for robotic applications such as
autonomous driving~\cite{kahn2018self,shah2023gnm} or nanocargo drug delivery~\cite{yang2021hierarchical,Feng2019}. 
One major difficulty in solving navigation tasks comes 
from the combined effects of the variability and complexity of the environment.
Variability can arise from hydrodynamic turbulence in animal navigation ~\cite{vergassola2007infotaxis,monthiller2022surfing,Reddy2016,calascibetta2023optimal,biferale2019zermelo,celani2014odor,Singh2023,Reddy2018},
and from thermal or non-equilibrium statistical fluctuations  
for active and driven colloids~\cite{colabrese2017flow,Nasiri2022,MuinosLandin2021SciRob,Yang2020,Pince2016,Grier2003Nature}
or bacterial chemotaxis~\cite{vladimirov2009chemotaxis}.
In addition, spatial complexity
makes optimal navigation policies non-trivial~\cite{Yang2018ACSNano,MuinosLandin2021SciRob,PiroTangGolestanian2021},
and its combination with fluctuations can lead to
transitions in the optimal policies~\cite{kappen2005path,Schneider2019,PiroMahaultGolestanian2022_NJP}.

Recently, advances in model-free Reinforcement learning 
have enabled the computation navigation policies
in intricate geometries such as mazes~\cite{Yang2018ACSNano} or complex energy landscapes~\cite{Nasiri2022,colabrese2017flow},
and in the presence of fluctuations, such as those generated by turbulent flow~\cite{Reddy2016,verano2023olfactory,Singh2023,rando2025q}
and thermal fluctuations~\cite{Boccardo2024PRE,MuinosLandin2021SciRob}.
In parallel, optimal policies can be computed by model-based
optimization methods. Such optimal policies not only help the fundamental understanding
of navigation, but also allow one to assess the performance of Reinforcement Learning policies~\cite{Heinonen2023,boccardo2022controlling,Boccardo2024PRE}.
 In this paper, we use model-based approaches to quantify the change in the optimal policies caused by disorder.
We demonstrate that introducing disorder via randomly distributed traps leads to a surprising non-monotonic effect: 
this change can peak at low trap concentrations and diminish as disorder increases.

We base our study on one of the most common Markov Decision Process models,
usually called gridworld in the Reinforcement Learning language~\cite{SuttonBarto2018},
where a navigation force biases a random walk on a lattice with traps.
For example, this could be achieved experimentally with colloids in optical lattices driven by hydrodynamic drag~\cite{Korda2002,Roichman2007},
or by laser-induced driving of colloids by asymmetric heating~\cite{MuinosLandin2021SciRob}.
The optimal policy 
is the optimal choice of the direction of the force in each site
that allows one to reach the target site
in minimum time. Using Dynamic Programming (DP)~\cite{SuttonBarto2018},
we compute the space-dependent distribution of optimal policies due to disorder.
To characterize the disorder-induced changes of the optimal policies we define
the \emph{density of change}, which is based on a Kullback-Leibler divergence.
This quantity has two interpretations, and therefore simultaneously answers two different questions:
(i) how does the optimal policy change when we add traps? or (ii) in a gridworld with traps,
how does the optimal policy change when we know where the traps are?
We compute spatial maps of the density of change, and find that it is non-monotonic
and exhibits a maximum when varying the trap concentration.

We then focus on the limit where the navigation bias is small compared
to the fluctuations, highlighting a regime of control fundamentally distinct from the strong-driving limit,
which is associated with minimal path problems~\cite{Buldyrev2004,Buldyrev2006,CordobaTorres2018,Alvarez2024,Villarrubia2024}
and deterministic optimal navigation~\cite{LiebchenLowen2019, PiroTangGolestanian2021,Zermelo1931navigationsproblem}.
In the small bias regime, the density of change is derived analytically
and exhibits a maximum at a low trap concentration that is inversely proportional to the trap strength.
This maximum, which persists at finite bias, does not
depend on how the navigation bias influences the transition rates, and should therefore pertain to 
a wide variety of navigation problems.
Finally, we show that finite size effects
can be described to leading order with the help
of the density of change caused by a single trap.

\paragraph{Model---}
Let us consider a Markovian continuous-time random walk on a two-dimensional square lattice, 
locally biased by a driving force $\mathbf{F}$ of fixed magnitude $F$.
This force can either be an internal force, e.g., created by a robot, or 
an external force applied by an external field.
The force orientation at each site $s$ is specified by a policy $\bm{\phi}$, such that $\mathbf{F}=F\,\bm{\phi}_s$.
The choice of the force orientation, referred to as the action, can take one of the four directions
of the first neighbors, 
such that for any site $s$, $\bm{\phi}_s \in \mathcal{A} = \{\pm\hat{\mathbf{x}},\, \pm\hat{\mathbf{y}}\}$.
During the dynamics, the state $s$ changes as a function of time,
and the policy $\bm{\phi}$ defines a feedback control process where 
the force is set in the course of time as a function of the current observed state $s$.
Our setting can therefore be seen as a stochastic and discrete version
of the Zermelo navigation problem~\cite{Zermelo1931navigationsproblem}.

Assuming thermally induced hops over barriers, as e.g. in diffusion of colloids in optical lattices~\cite{Korda2002,Evstigneev2008,Brazda2018}, 
or at the surface of colloidal crystals~\cite{Mondal2020}, the transition rate from site $s$ to a neighboring site $s' \in\mathcal{B}_s$ reads ~\cite{Hanggi1990RMP,Kramers1940Physica}
\begin{align}
\label{eq:rates_model}
\gamma^{\bm{\phi}}_{s's}
&=
\gamma^{0}_{s}
\exp\!\Bigl[
F\bm{\phi}_s\!\cdot\!\mathbf{u}_{s's}/k_{\mathrm{B}}T
\Bigr],
\end{align}
where $\gamma^{0}_{s}=\nu\exp[-E^{0}_{s}/(k_{\mathrm{B}}T)]$ 
denotes the rate when $F=0$, with
$\nu$ an attempt frequency and
$E^{0}_{s}$ the diffusion barrier, 
$\mathbf{u}_{s's}$ is a vector of length $d/2$ 
pointing from $s$ to $s'$ with $d$ the lattice constant, 
and $k_{\mathrm{B}}T$ is the thermal energy.

\begin{figure}[ht]
    \centering
    \includegraphics[width=\columnwidth]{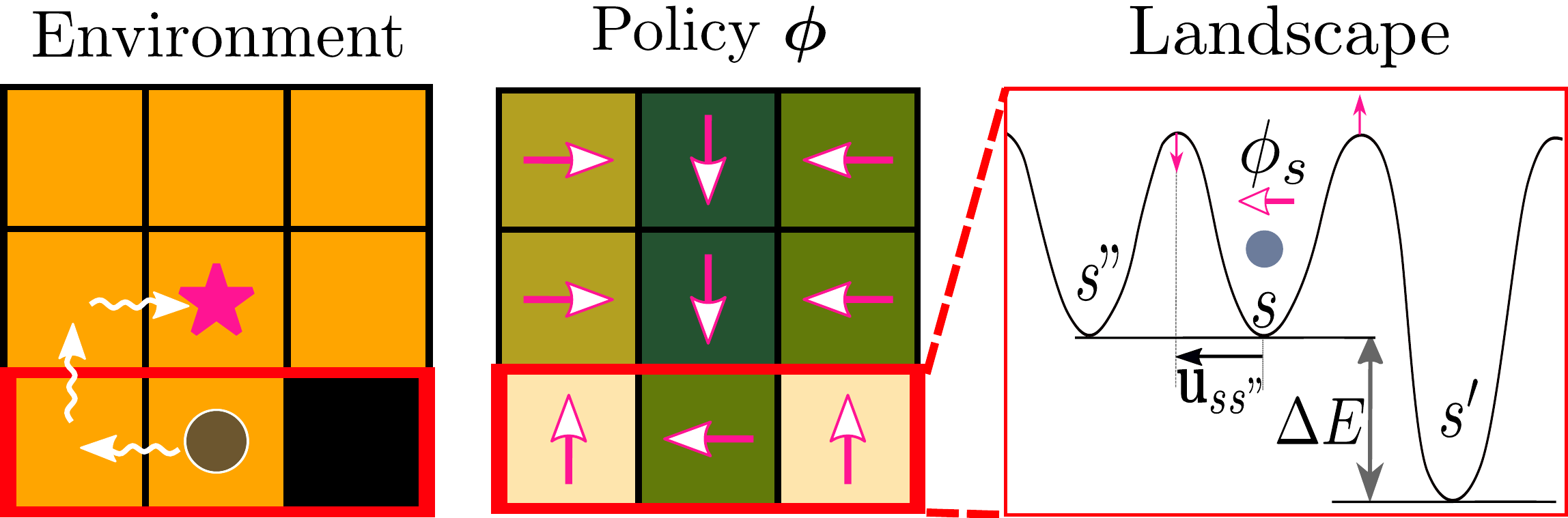}
    \caption{
    Gridworld navigation model. Left: a particle diffuses,
    and reaches the target state indicated by the star.  Center: 
    the policy $\bm{\phi}$ is defined in each state.
    Right: we use a simple barrier-passing model. The policy
    increases the rate along the force
    and decreases them in the opposite direction. Black sites
    are traps with a deeper energy well.
    }
    \label{fig:model}
\end{figure}

In the homogeneous case, all barriers are identical $E^{0}_{s}=E^{0\mathrm{h}}$, 
so that $\gamma^{0}_{s}=\gamma^{0\mathrm{h}}$ is independent of $s$. 
Quenched disorder is modeled by a Random Trap Model~\cite{Bouchaud1992JPI,MonthusBouchaud1996PRE,BouchaudGeorges1990PhysRep}, 
where trap sites have a deeper potential well, corresponding to a larger barrier $E^{0}_{s}=E^{0\mathrm{h}}+\Delta E$, 
with $\Delta E>0$, as shown in \cref{fig:model}. 
The rates for escaping from traps are thus decreased by a factor $R=\exp(\Delta E/k_{\mathrm{B}}T) > 1$,
called the trap strength.

\paragraph{Optimal policies--- }   
We define the mean first-passage time (MFPT) $T^{\bm{\phi}}_{\bar{s}s}$ 
to reach the target $\bar{s}$ starting from site $s$,
and following the  policy $\bm{\phi}$. 
Our goal is to find an optimal policy $\bm{\phi}^\star$ that minimizes $T^{\bm{\phi}}_{\bar{s}s}$ for all sites $s$, 
leading to $T^{\star}_{\bar{s}s}=\min_{\bm{\phi}} T^{\bm{\phi}}_{\bar{s}s}$. 
This optimization problem is a Markov decision process, and the optimal MFPT satisfies the Bellman optimality equation~\cite{SuttonBarto2018}
\begin{align}
\label{eqn:Bellman_optimality_equation}
T_{\bar{s}s}^{\star}
&=
\min_{\bm{\phi}_s \in \mathcal{A}}
\Bigl[
t^{\bm{\phi}}_{s}
+
\sum_{s' \in \mathcal{B}_s}
p_{s's}^{\bm{\phi}}\, T_{\bar{s}s'}^{\star}
\Bigr],
\end{align}
where $T_{\bar{s}\bar{s}}^{\star}=0$ at the target,
$t^{\bm{\phi}}_{s}
=
1/\bigl(\sum_{s' \in\mathcal{B}_s}\gamma^{\bm{\phi}}_{s's}\bigr)$ are the average residence times and 
$p^{\bm{\phi}}_{s's}=\gamma^{\bm{\phi}}_{s's}t^{\bm{\phi}}_{s}$ are the transition probabilities.
The optimal policy is found numerically using DP:
we use an iterative scheme based on \cref{eqn:Bellman_optimality_equation} called 
value iteration~\cite{SuttonBarto2018}. Moreover, we consider reflective boundaries.

\begin{figure}[ht]
    \centering
    \includegraphics[width=0.9\columnwidth]{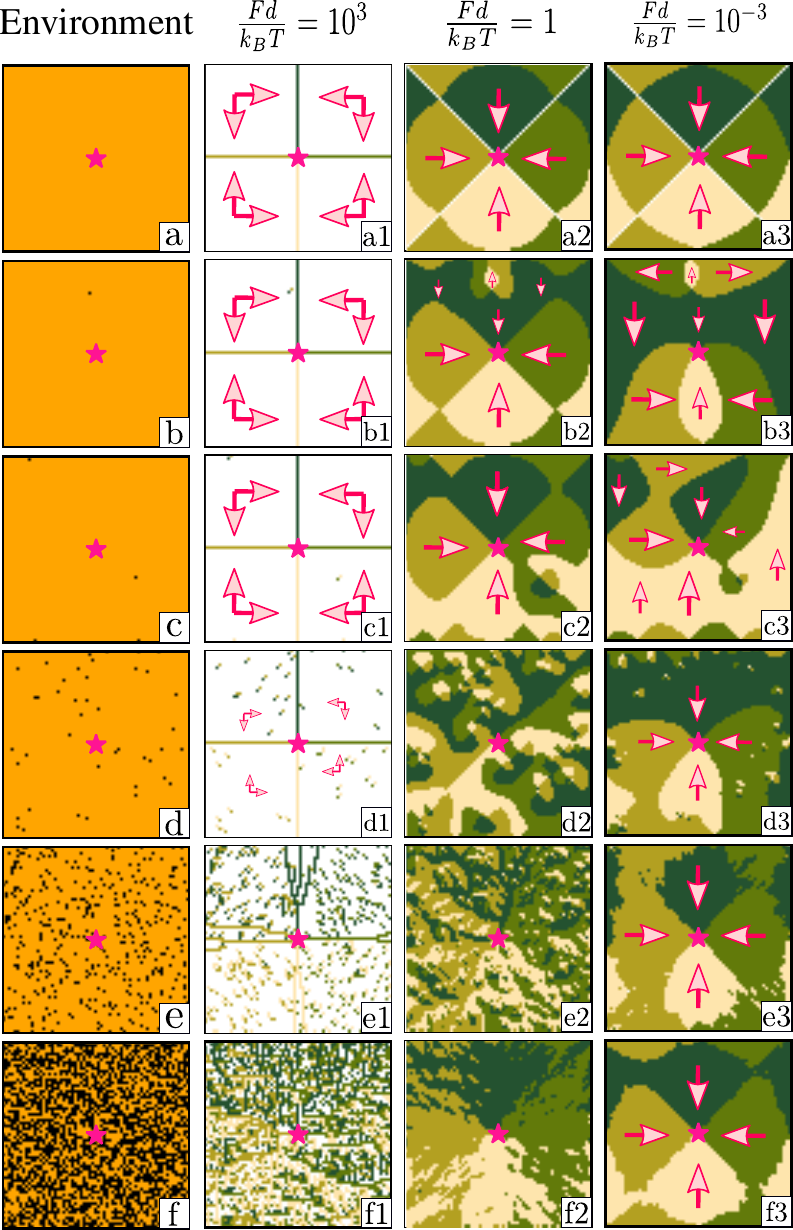}
    \caption{
    {Optimal policies in gridworld with traps.}
    \label{fig:policy_evol}
    Results obtained via DP on a $65\times65$ lattice.
    Trap depth: $\Delta E/k_BT = 10$.
    Colored regions denote sites with a unique optimal orientation; white regions indicate a degenerate policy.
    (a) Homogeneous environment.
    (b) Single trap.
    (c) Four traps.
    (d) $1\%$ trap concentration $c_d=0.01$.
    (e) $10\%$, $c_d=0.1$.
    (f) $50\%$, $c_d=0.5$.
    }
\end{figure}

Solving \cref{eqn:Bellman_optimality_equation} with DP for different realizations 
of the energy landscape provides the optimal policies shown in \cref{fig:policy_evol}. 
Dynamic Programming theory stipulates that while optimal MFPT $T_{\bar{s}s}^{\star}$ are unique, 
the optimal policy may not be~\cite{SuttonBarto2018}.
Indeed, more than one action can be optimal at a given site, we will say that the policy is degenerate at this site.
Degenerate sites are shown in white in \cref{fig:policy_evol}.
While exact degeneracy arises from symmetries of the problem,
approximate degeneracy reflects numerically indistinguishable MFPT values for distinct actions 
(see numerical methods in SM).

In a homogeneous lattice [\cref{fig:policy_evol}(a)], the optimal policy depends only on the dimensionless bias strength $Fd/k_{\mathrm{B}}T$. 
In the large-force limit, $Fd/k_{\mathrm{B}}T \gg 1$, transitions occur predominantly along the force orientation $\bm{\phi}_s$.
Hence, the minimal MFPT is achieved by a policy pointing towards the shortest path to the target, 
which corresponds to Manhattan geodesics on the square lattice \cite{kardar2007, Newman2010,CordobaTorres2018}.
The associated policy is approximately degenerate in the 4 quadrants of the lattice as shown in \cref{fig:policy_evol}(a1).
In contrast, in the weak-force regime, $Fd/k_{\mathrm{B}}T \ll 1$, thermal fluctuations dominate and 
 $\bm{\phi}_s$ only leads to a small increase of the transition rate along the force direction.
Moreover, the diagonals of the lattice are seen to exhibit exact degeneracy by symmetry in \cref{fig:policy_evol}(a).

When the trap concentration $c_d$ is small,
and in the large-force regime $Fd/k_{\mathrm{B}}T \gg 1$, as in \cref{fig:policy_evol}(b1-d1),
the influence of each trap remains spatially localized,
and is restricted to removing the degeneracy around the defects to avoid them. 
Interestingly, tree-like non-degenerate regions can be observed
at intermediate trap concentrations in \cref{fig:policy_evol}(e1).
The largest changes of the policy arise at finite concentrations. 
In contrast, the weak-force regime $Fd/k_{\mathrm{B}}T \ll 1$ exhibits a striking sensitivity to traps. 
Even a single trap in \cref{fig:policy_evol}(b3) induces a global change in the optimal policy.
As the concentration $c_d$ of defects increases [\cref{fig:policy_evol}(c3–f3)], the policy 
first changes more and more, and then
gradually comes back towards that of the homogeneous case. 
As opposed to the large force regime, the largest deviations of the optimal policy from the homogeneous case occur at low defect densities.
This behavior persists up to finite bias, when $Fd/k_{\mathrm{B}}T\sim 1$,
as seen from \cref{fig:policy_evol}(b2–f2).

\noindent
\paragraph{Density of change---}
We now aim to provide a quantitative description
of the changes in the policy caused by disorder.
In the following, instead of the
deterministic optimal policies $\boldsymbol{\phi}^\star$, it is convenient to define  probabilistic optimal policies
by assigning equal probability to each degenerate action
$\pi^\star_{\bar{s}s}(\bm{a})=\mathbb{I}_{a\in {\cal A}_{\bar{s}s}^\star}/|{\cal A}_{\bar{s}s}^\star|$, 
where ${\cal A}_{\bar{s}s}^\star$ is the set of optimal actions at site $s$,
$|{\cal A}_{\bar{s}s}^\star|$ is the number of optimal actions at $s$,
and $\mathbb{I}$ is the indicator function.
To each 
realization of disorder,
we associate a probabilistic optimal policy $\pi^{\star\mathrm{d}}_{\bar{s}s}(\bm{a})$.
Our goal is to characterize the distribution of these policies,
and our main focus will be on their average over disorder $\langle \pi^{\star\mathrm{d}}_{\bar{s}s}\rangle(\bm{a})$.
To quantify how  $\langle \pi^{\star\mathrm{d}}_{\bar{s}s}\rangle$ differs from the optimal policy
in a homogeneous system $\pi^{\star \mathrm{h}}_{\bar{s}s}$,
we introduce the \emph{local density of change} $\rho_{\bar{s}s}$.
For a site $s$ with a non-degenerate policy in the homogeneous environment, i.e. with a unique optimal action 
$\bm{a}=\phi^{\star\mathrm{h}}_s$, 
we define $\rho_{\bar{s}s}$ as the probability that the optimal action 
in a disordered environment differs from that of the homogeneous environment
\begin{align}
\label{eqn:non-degenerate_site_DoC}
\rho_{\bar{s}s}
&=
1 - \langle \pi^{\star\mathrm{d}}_{\bar{s}s}\rangle(\boldsymbol{\phi}^{\star\mathrm{h}}_s).
\end{align}
To extend the definition of $\rho_{\bar{s}s}$ to degenerate sites, we require that $\rho_{\bar{s}s}=0$ if and only if 
$\langle \pi^{\star\mathrm{d}}_{\bar{s}s}\rangle(\bm{a})=\pi^{\star\mathrm{h}}_{\bar{s}s}(\bm{a})$
for all $\bm{a}$. A definition that satisfies this constraint and reduces to 
\cref{eqn:non-degenerate_site_DoC} in the non-degenerate case is
(see SM for detailed derivations)
\begin{align}
\label{eqn:Relation_KL_DoC}
\rho_{\bar{s}s}
&=
1-\exp\!\left[
-\mathcal{D}_{\bar{s}s}
\bigl[\pi^{\star\mathrm{h}}_{\bar{s}s}\|
\langle \pi^{\star\mathrm{d}}_{\bar{s}s}\rangle\bigr]
\right],
\end{align}
where
$\mathcal{D}_{\bar{s}s}[\pi_1\|\pi_2]=\sum_{\bm{a}\in \mathcal{A}}\pi_1(\bm{a})\ln[\pi_1(\bm{a})/\pi_2(\bm{a})]$
is the Kullback–Leibler divergence between $\pi_1$ and $\pi_2$. 
Since the  Kullback–Leibler divergence is positive, we have $0\leq\rho_{\bar{s}s}\leq 1$.

\begin{figure*}[ht] 
    \centering 
    \includegraphics[width=\linewidth]
    {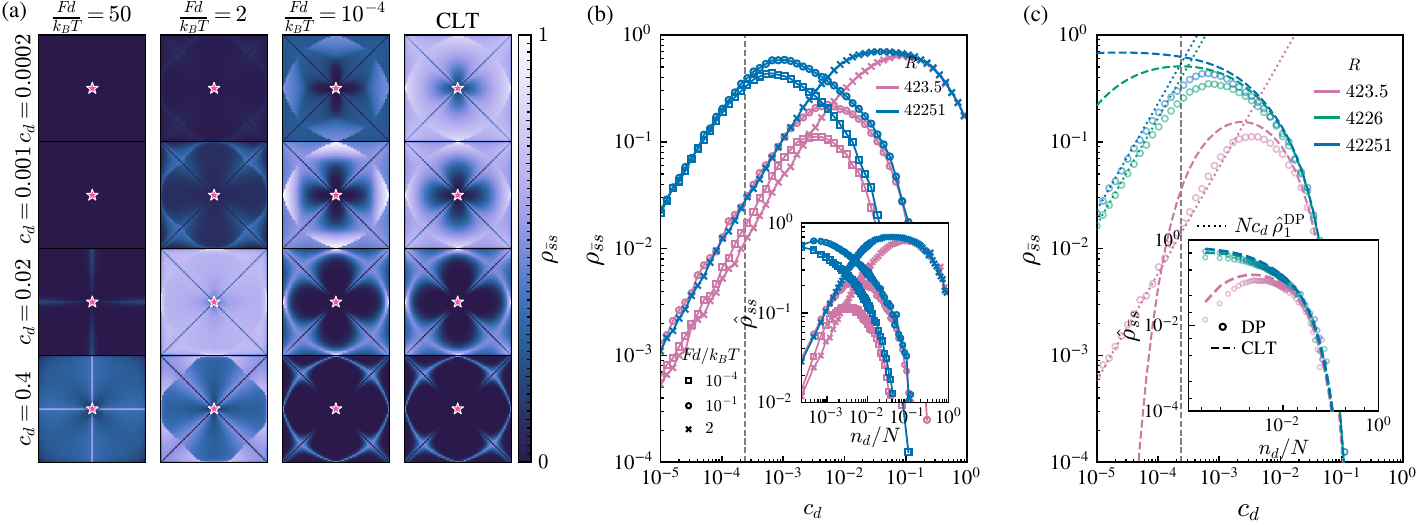}
    \caption{{Density of change of the optimal policy.}
    (a) Maps computed via DP on a $65 \times 65$ square lattice at $R = 42251$.
    The rightmost column reports the analytical results of the CLT.
    (b) DP simulation in  a $65 \times 65$ gridworld at $s=(8,20)$.
    (c) Weak force regime at  $s=(8,20)$ with $Fd/k_{\mathrm{B}}T=10^{-4}$, and CLT prediction.
    In (b) and (c), the insets show the local density of change $\hat{\rho}_{\bar{s}s}$ at fixed number of traps $n_d$
    extracted from the same DP data.
    In (a,b,c), each point is obtained from an average over $2000$ disorder realizations with at least one defect. 
    }
    \label{fig:DP_density_of_change}
\end{figure*}

We assume that the traps are independently distributed at each lattice site according to a Bernoulli law
with an average concentration $c_d$.
Averaging DP policies over disorder realizations at fixed $c_d$, 
we obtain maps of $\rho_{\bar{s}s}$ shown in \cref{fig:DP_density_of_change}(a). 
At large forces, $\rho_{\bar{s}s}$ is maximum along the $x$ and $y$ axes
passing through the target, where the policy of the homogeneous system $\pi^{\star\mathrm{h}}$ was
non-degenerate.
At small forces, the maps are qualitatively different, 
$\rho_{\bar{s}s}$ is low close to the target where the policy does not change because it  mostly stays directed toward the target,
and on the diagonals where $\pi^{\star\mathrm{h}}_{\bar{s}s}$ and
$\langle \pi^{\star\mathrm{d}}_{\bar{s}s}\rangle$ are both  degenerate by symmetry.

In addition, \cref{fig:DP_density_of_change}(a) provides
a quantitative assessment of the non-monotonicity of $\rho_{\bar{s}s}$ when varying $c_d$
at low and moderate forces. This is confirmed by the evolution of $\rho_{\bar{s}s}$
at a given point of the lattice as a function of $c_d$ in \cref{fig:DP_density_of_change}(b).

\paragraph{Weak-force expansion---}
 We now focus on computing $\rho_{\bar{s}s}$ in the weak force regime,
and finding the concentration of defects for which $\rho_{\bar{s}s}$ is maximal.
In the regime $Fd/k_\mathrm{B}T \ll 1$, the transition rates can be linearized as
\begin{align}
\label{eqn:small_force_rates_expansion}
\gamma^{\bm{\phi}\mathrm{p}}_{s's}
&= \gamma^{0\mathrm{p}}_{s}
+ \frac{F}{k_\mathrm{B}T}\,
\bm{\phi}_{s}\!\cdot\!\mathbf{u}_{s's}\,\gamma^{0\mathrm{p}}_{s}
+ \mathcal{O}\!\left[(Fd/k_{\mathrm{B}}T)^2\right],
\end{align}
where the superscript $0$ refers to the zero-force case ($F=0$),
and $\mathrm{p}=\mathrm{h}$ or $\mathrm{d}$ respectively refer to the homogeneous case or
a realization of disorder.
When $Fd/k_\mathrm{B}T \ll 1$, the optimal policy is independent of
$F$.
It is along the direction that decreases the most the
MFPT without force in the sense that it maximizes the projection
on the opposite of the gradient of the MFPT
(see~\cite{Boccardo2024PRE} and SM)
\begin{align}
\label{eqn:optimal_policy_small_force}
\bm{\phi}^{\star\mathrm{p}}_{s}
& \in \argmax_{\bm{\phi}_{s} \in \mathcal{A}}
\bigl\{
-\mathbf G^{\mathrm{p}}_s
\cdot \bm{\phi}_{s}
\bigr\},
\end{align}
where 
$\mathbf G^{\mathrm{p}}_s= \nabla^\dagger_{\gamma^{0\mathrm{p}}_s}\, T^{0\mathrm{p}}_{\bar{s}s}$, 
and the gradient of a scalar $v_s$ reads
\begin{align}
\label{eqn:gradient_definition}
\nabla^\dagger_{\gamma^{\boldsymbol{\phi}\mathrm{p}}_s}\, v_s
&\equiv
\sum_{s' \in \mathcal{B}_s}
\gamma^{\boldsymbol{\phi}\mathrm{p}}_{s's}\,
\bigl(v_{s'} - v_{s}\bigr)\,\mathbf{u}_{s's}.
\end{align}

We now use the MFPT decomposition $T^{0\mathrm{p}}_{\bar{s}s}=\sum_{s'}\Xi^{0 \mathrm{p}}_{s'\bar{s}s}$
into occupation times $\Xi^{0 \mathrm{p}}_{s'\bar{s}s}$, defined as the total time spent at site $s'$  starting from $s$ before reaching $\bar{s}$ for the first time~\cite{AldousFill,Benichou2014PhysRep}. 
In the Random Trap Model, we have 
$\Xi^{0 \mathrm{d}}_{s'\bar{s}s}=(\gamma^{0 \mathrm{h}}_{s'}/\gamma^{0 \mathrm{d}}_{s'})\,\Xi^{0 \mathrm{h}}_{s'\bar{s}s}$,
where $\Xi^{0 \mathrm{h}}_{s'\bar{s}s}$ can be computed exactly~\cite{Benichou2014PhysRep}.

Using these relations, the gradient needed to compute the optimal policy through \cref{eqn:optimal_policy_small_force} is written as
\begin{equation}
\label{eqn:Gradient_disorder_with_Theta}
\mathbf G^{\mathrm{d}}_s
=
\bm{\Gamma}_{s\bar s s}
+
\frac{\gamma^{0 \mathrm{d}}_{s}}{\gamma^{0 \mathrm{h}}_{s}}\,\bm{\Theta}_{\bar s s},
\end{equation}
where we have defined
\begin{equation}
\nonumber
\bm{\Gamma}_{s'\bar s s}
=
\nabla^\dagger_{\gamma^{0 \mathrm{h}}_s}
\Xi^{0 \mathrm{h}}_{s'\bar s s}, 
\qquad
\bm{\Theta}_{\bar s s}
=
\sum_{s'\neq s}
\frac{\gamma^{0 \mathrm{h}}_{s'}}{\gamma^{0 \mathrm{d}}_{s'}}\bm{\Gamma}_{s'\bar s s}.
\end{equation}

We denote by a subscript $[i]$ quantities conditioned on the presence of a trap at a given site $s$, where $i=1$ if $s$ is a trap and $i=0$ otherwise. 
Under the Central Limit approximation, we compute  $\mathcal Q_{[i]}(\bm{a})$, 
the conditional probability that action $\bm{a}$ satisfies \cref{eqn:optimal_policy_small_force}.
The details of this calculation are reported in SM.
To align the boundaries between two optimal actions defined 
by \cref{eqn:optimal_policy_small_force} with the coordinate axes,
we use the basis $(\hat{\mathbf u},\hat{\mathbf v})$ 
rotated by $\pi/4$ with respect to $(\hat{\mathbf x},\hat{\mathbf y})$. 

We then find
\begin{align}
\label{eqn:Q_i_closed_form}
\mathcal{Q}_{[i]}(\bm{a})
&=
1 
-\!\!\!\sum_{\mathbf{k}\in\{\mathbf{u},\mathbf{v}\}}\!\!\!\Phi\bigl(\alpha^{\mathbf{k}}_{[i]}(\bm{a})\bigr)
+\Phi_2\bigl({\boldsymbol \alpha}_{[i]}(\bm{a});\kappa(\bm{a})\bigr),
\end{align}
where 
$X^{\mathbf k}=\mathbf X\cdot{\mathbf k}$, with $\mathbf{k}=\mathbf{u}$, or $\mathbf{v}$
denote the components of the vector $\mathbf X$,
$\alpha^{\mathbf{k}}_{[i]}(\bm{a})
={\sgn(a^\mathbf{k})\,m^{\mathbf{k}}_{[i]}}/{\sigma^{\mathbf{k}}_{[i]}}$,
with
\begin{align}
\nonumber
\bm m_{[i]}
&=
\bm{\Gamma}_{s\bar{s}s}
+
(c_dR^{1-i}+(1-c_d)R^{-i})
\sum_{s'\neq s} \bm{\Gamma}_{s'\bar{s}s},
\\
\sigma^{\mathbf{k}}_{[i]}
&=
R^{-i}[c_d(1-c_d)(R-1)^2
\sum_{s' \neq s}
\bigl(\Gamma^{\mathbf{k}}_{s'\bar{s}s}\bigr)^2]^{1/2},
\nonumber\\
\kappa(\bm{a})&= \sum_{s'\neq s}
\prod_{\mathbf{k}\in\{\mathbf{u},\mathbf{v}\}}
\frac{\sgn(a^\mathbf{k} )\Gamma_{s'\bar s s}^{\mathbf k} }
{[\sum_{s' \neq s}(\Gamma_{s'\bar{s}s}^{\mathbf{k}})^2]^{1/2}},
\nonumber
\end{align}
and the cumulative distributions $\Phi(x)=\mathrm{erfc}[-x/2^{1/2}]/2$ and 
$\Phi_2(\mathbf{x};\kappa)
=
\int_{-\infty}^{x^{\mathbf{u}}}\hspace{-0.1 cm}\mathrm{d}\mu\int_{-\infty}^{x^{\mathbf{v}}}\hspace{-0.1 cm}\mathrm{d}\nu
\,\mathrm{e}^{-\chi/2\kappa_1^2}/(2\pi \kappa_1)$, with $\chi=\mu^2-2\kappa \mu\nu+\nu^2$
and $\kappa_1=(1-\kappa^2)^{1/2}$.

Once $\mathcal{Q}_{[i]}(\bm{a})$ is known, we obtain the disorder-averaged optimal-action probability as
\begin{align}
\label{eqn:pi_dis_Q}
\langle \pi^{\star\mathrm{d}}_{\bar{s}s}\rangle(\bm{a})
=
(1-c_d)\,\mathcal{Q}_{[0]}(\bm{a})
+
c_d\,\mathcal{Q}_{[1]}(\bm{a}).
\end{align}

The CLT prediction for $\rho_{\bar{s}s}$ combining \cref{eqn:pi_dis_Q,eqn:Relation_KL_DoC} is in quantitative agreement with DP simulations for finite $c_d$, 
as shown in \cref{fig:DP_density_of_change}(a,c) (see SM Fig.S6 for maps of $\langle \pi^{\star\mathrm{d}}_{\bar{s}s}\rangle$).
Detailed quantitative agreement in the whole simulation box is shown in SM Fig.S5.
Since the CLT is based on the limit $N\rightarrow\infty$ for fixed values of $c_d$ and $R$,
it assumes a large number of defects $n_d\approx N c_d\gg 1$. In finite systems, like our
$65\times65$ gridworld investigated with DP, the CLT prediction 
deviates from DP when $n_d$ is small, as seen in \cref{fig:DP_density_of_change}(a,c).

In order to probe the finite-size regime, 
we consider the very dilute regime $c_d \ll 1/N$, where disorder realizations contain at most one trap.
In this regime,  $\rho_{\bar{s}s}(c_d)\approx Nc_d \,\hat\rho_{\bar{s}s}(1)$,
where $\hat\rho_{\bar{s}s}(n_d)$ is the density of change at fixed number
of traps $n_d$. This relation is confirmed
in \cref{fig:DP_density_of_change}(c) with $\hat\rho_{\bar{s}s}(1)$
extracted from DP.
An approximate expression for the one-defect density of change $\hat\rho_{\bar{s}s}(1)$
in the large volume limit~\cite{Benichou2014PhysRep} for a single defect 
is provided in SM.

One can also extract $\langle\pi^{\star \mathrm{d}}_{\bar{s}s}\rangle$
as a function of the exact number of traps $n_d$ from DP,
as reported in the insets of \cref{fig:DP_density_of_change}(b,c).
The corresponding prediction from the CLT for the density of change  
is shown in \cref{fig:DP_density_of_change}(c) using the approximation $\hat\rho_{\bar{s}s}(n_d)\approx \rho_{\bar{s}s}(c_d=n_d/N)$.

Using  Eqs.(\ref{eqn:pi_dis_Q}) and (\ref{eqn:Relation_KL_DoC}), 
the maximum of $\rho_{\bar{s}s}(c_d)$ is predicted to be inversely proportional to the 
trap strength (see SM)
\begin{align}
c_d^\mathrm{max}= \frac{1}{R+1}.
\label{eq:cdmax}
\end{align}
Note that $c_d^\mathrm{max}$  does not depend on the system size $N$
in this CLT prediction. In finite systems, since $\rho_{\bar{s}s}$ increases linearly at small $c_d$,
we expect $\rho_{\bar{s}s}$ to be maximum at $c_d\approx \max[c_d^\mathrm{max},1/N]$.
Hence, the prediction \cref{eq:cdmax}  holds for $R\sim 1/c_d^\mathrm{max} <N$,
in agreement with the results reported in \cref{fig:DP_density_of_change}(c)
with $N=4225$.

Three additional remarks are worth noting to examine the  significance of our results.
First, the non-monotonic behavior of $\rho_{\bar{s}s}$ with trap concentration
is generic in the sense that it does not rely on the specific expression of the transition rates \cref{eq:rates_model}.
The existence of the maximum and \cref{eq:cdmax}  extends to any model where the transition rates 
can be linearized in the driving force,
i.e. $\gamma^{\boldsymbol{\phi}}_{s's}=\gamma_s^0+F\boldsymbol{\phi}_s\cdot \boldsymbol{\psi}_{s's}+O(F^2)$.
As a consequence, our results should apply not only to driven colloidal navigation systems such as those driven by hydrodynamic drag~\cite{Korda2002,Roichman2007},
that share strong similarities with our model, 
but also to experiments with different physics for the transition rates, 
as in  laser-induced driving by asymmetric heating~\cite{MuinosLandin2021SciRob}.

Second remark, since a given action $\bm{a}$ at site $s$
either coincides or differs from
the optimal policy $\bm{\phi}^\mathrm{\star d}_{s}$ 
in a given realization of disorder, we conclude that
$\mathbb{I}_{\bm{a}=\bm{\phi}^\mathrm{\star d}_{s}}$ is a Bernoulli random variable~\footnote{
This statement assumes a random choice of the actions with equal probability among degenerate optimal actions
for each realization of disorder.}. 
Thus, from its average $\langle\pi^\mathrm{\star d}_{\bar{s}s}\rangle(\bm{a})$,
the full policy distribution can be computed~\cite{bertsekas2002introduction}. 
We report maps of the policy variance in SM Fig.S7. 

The final remark pertains to the case of an agent
not knowing the position of the traps. 
Since the mean residence times averaged over disorder are simply multiplied by a factor 
independent of space $\langle t_s^{\boldsymbol{\phi}\mathrm{d}}\rangle=(1-c_d+Rc_d)t_s^{\boldsymbol{\phi}\mathrm{h}}$,
while the transition probabilities do not change $\langle p_{s's}^{\boldsymbol{\phi}\mathrm{d}}\rangle=p_{s's}^{\boldsymbol{\phi}\mathrm{h}}$,
the optimal policy $\pi^{\star\mathrm{b}}$ for such an agent is the same as that of the homogeneous environment $\pi^{\star\mathrm{b}}=\pi^{\star\mathrm{h}}$.
This leads to an alternative interpretation of our results:
the density of change $\rho_{\bar{s}s}$ also quantifies the change in the policy
due to the knowledge of the trap positions.

\paragraph{Conclusion---}
We analyzed the optimal navigation for a Brownian particle in a random-trap landscape. 
Upon the decrease of the ratio between the strength of the driving force and the 
strength of fluctuations, the policy changes from
the direction of a time-weighted path length minimization,
to the direction that decreases the most the MFPT without force.
Our simple minimal gridworld model points to a non-monotonic change
of the optimal policy at small trap density for small and moderate biases. 
This behavior is generic in the sense that
it does not depend on the specific dependence of the rates on the actions,
and could therefore be relevant not only for navigation experiments with colloids,
but also for other navigation problems,
including those involving animals and robots.

\bibliographystyle{apsrev4-2}
\bibliography{references}

\end{document}